\documentclass{aa}  
\usepackage{amsmath}
\usepackage{graphicx}
\usepackage{txfonts}
\usepackage{multirow}
\begin{document}

\title{A $\gamma$-ray emitting NLS1 galaxy SDSS J095909.51+460014.3 }

\subtitle{identified by multiwavelength contemporaneous brightening}

\author{Yang-Ji Li
          \inst{1}
          \and
             Neng-Hui Liao
           \inst{1} 
           %\thanks{nhliao@gzu.edu.cn}
           \and
           Zhen-feng Sheng 
           \inst{2}  \fnmsep \inst{4}
           %\thanks{zfsheng@gzu.edu.cn}
           \and
           Sina Chen
           \inst{5} 
           \and
           Yi-Bo Wang
          \inst{3} \fnmsep \inst{4} 
           \and
           Ting-Gui Wang 
          \inst{3} \fnmsep \inst{4}
          }

\institute{Department of Physics and Astronomy, College of Physics, Guizhou University, Guiyang 550025, China, \email{nhliao@gzu.edu.cn};\
         \and
             Institute of Deep Space Sciences, Deep Space Exploration Laboratory, Hefei 230026, China, \email{shengzf@ustc.edu.cn};
         \and
            Deep Space Exploration Laboratory / Department of Astronomy, University of Science and Technology of China, Hefei 230026, China
         \and
            School of Astronomy and Space Science, University of Science and Technology of China, Hefei 230026, China\
          \and
           Department of Physics, Technion Israel Institute of Technology,  Haifa 3200003, Israel }

   \date{Received xxx, 2022; accepted xxx, 2023}

 \abstract{We report on an identification of a new $\gamma$-ray emitting narrow-line Seyfert 1 galaxy ($\gamma$-NLS1), SDSS J095909.51+460014.3 ($z$ = 0.399), by establishing an association with a $\gamma$-ray source 4FGL 0959.6+4606, although its low-energy counterpart was suggested to be a radio galaxy 2MASX J09591976+4603515. \emph{WISE} long-term light curves of these two sources reveal diverse infrared variability patterns. Brightenings of 2.5 mag are detected for the former source, while flux decays of 0.5 mag are found for the other one. More importantly, the time that the infrared flux of the NLS1 rises, is coincident with the time of flux increase of 4FGL 0959.6+4606. At the same time, no infrared activity of the radio galaxy has been observed. A specific analysis of 15-month Fermi-LAT data, aiming at the high $\gamma$-ray flux state, yields a significant source (TS =43). The corresponding $\gamma$-ray localization analysis suggests that only the NLS1 falls into the uncertainty area, further supporting the updated association relationship. A broadband spectral energy distribution of SDSS J095909.51+460014.3 has been drawn and well described by the classic single-zone homogeneous leptonic jet model. Its jet properties are investigated and found to be comparable with the known $\gamma$-NLS1s.}
% 5 {} token are mandatory
 
%  \abstract
%   {xxx}
%   {xxx}
%   {xxx}
%   {xxx}
\maketitle
%
%________________________________________________________________
\section{Introduction} \label{sec:intro}
In the extragalactic $\gamma$-ray sky, active galactic nuclei (AGNs) with strong jets powered by accretion of materials onto supermassive black holes (SMBHs) act as the dominate population \citep{2016ARA&A..54..725M}. Most of these sources are blazars, including flat-spectrum radio quasars (FSRQs) and BL Lacertae objects (BL Lacs). Their relativistic jets are closely aligned with our line of sight and hence the Doppler boosted, highly variable jet emissions are overwhelming \citep{1978bllo.conf..328B,1997ARA&A..35..445U,2019ARA&A..57..467B}. The jet emissions are featured as a universal two-bump structure in log$\nu$F$\nu$-log$\nu$ plot, in which one is due to synchrotron emission while the other one enters into the $\gamma$-ray domain. In the leptonic scenarios, the latter is explained as inverse Compton (IC) scattering of soft photons, either inside (synchrotron self-Compton, or SSC, \citealt{1992ApJ...397L...5M}) and/or outside of the jet (external Compton, or EC, \citealt{1993ApJ...416..458D,1994ApJ...421..153S,2000ApJ...545..107B}). On the other hand, hadronic scenarios are adopted to describe blazar $\gamma$-ray flares temporally coincident with the cospatial neutrino detection \citep[e.g.,][]{2018Sci...361.1378I,2020ApJ...893..162F,2022ApJ...932L..25L}.

In addition to blazars, Fermi-LAT observations reveal that radio-loud narrow-line Seyfert 1 (RLNLS1) is a new subclass of $\gamma$-ray emitting AGN \citep{2009ApJ...699..976A,2009ApJ...707L.142A}.  NLS1s are characterized by an optical spectrum with narrow permitted lines (i.e. FWHM~(H$\beta$) < 2000 km/s), weak [O\,\textsc{iii}] line emission (i.e. [O\,\textsc{iii}]/H$\beta$ < 3), and strong optical Fe\,\textsc{ii} emission \citep{2000NewAR..44..381P}. Prominent soft X-ray excesses are also exhibited \citep{1996A&A...309...81W,1996A&A...305...53B}. Only a small fraction of NLS1s harbor strong jets (i.e. radio loudness $\mathcal{R} = f_{5~GHz}/f_{B} > 10$, \citealt{1989AJ.....98.1195K,2006AJ....132..531K}). Evidences of the presence of relativistic jets in RLNLS1s have been noticed, including compact morphology, flat or inverted spectral slopes, a very high brightness temperature and significant variability \citep{2003ApJ...584..147Z,2007ApJ...658L..13Z,2008ApJ...685..801Y}. Detections of $\gamma$-ray emissions of RLNLS1s then provide a decisive proof that their central engines resemble that of blazars. However, compared with blazars, RLNLS1s tend to possess accretion systems with relatively under-massive SMBHs in a high Eddington ratio (\citealt{2002ApJ...565...78B}, but also see \citealt{2008MNRAS.386L..15D,2016MNRAS.458L..69B}). Meanwhile, host galaxies of NLS1s are suggested to be in an early phase of galaxy evolution \citep{2007ApJS..169....1O}. But blazars are usually hosted in giant elliptical galaxies (\citealt{2000ApJ...543L.111L,2007ApJ...658..815S}, but also see \citealt{2011A&A...535A..97M}). The diverse properties of the accretion and the host environments between RLNLS1s and blazars make $\gamma$-ray emitting RLNLS1s (i.e. $\gamma$-NLS1s) valuable targets, shedding light on the formation and evolution of relativistic jets under extreme physical conditions as well as the coupling of jets and accretion flows.
 
Among several thousand extragalactic $\gamma$-ray sources \citep{2020ApJ...892..105A}, there are only a handful $\gamma$-NLS1s \citep{2009ApJ...699..976A,2009ApJ...707L.142A,2012MNRAS.426..317D,2015MNRAS.452..520D,2015MNRAS.454L..16Y,2018MNRAS.477.5127Y,2018ApJ...853L...2P,2019MNRAS.487L..40Y,2021MNRAS.504L..22R}. These sources have drawn great attentions and triggered extensive multiwavelength studies \citep[e.g.,][]{2012ApJ...759L..31J,2015ApJS..221....3G,2015A&A...575A..13F,2019A&A...632A.120B}. Considering the relatively limited angular resolution of Fermi-LAT \citep{2009ApJ...697.1071A}, multi-band correlated activities are needed to pin down the association between the $\gamma$-ray source and the low-energy counterpart. In such a case, the number is even smaller.

4FGL 0959.6+4606 is included in the latest 4FGL-DR3 catalog \citep{2022ApJS..260...53A}. 2MASX J09591976+4603515 ( hereinafter candidate A), an edge-on radio galaxy (RG, $z$ = 0.148, \citealt{2021AJ....162..177P}), is listed as its low-energy counterpart. Candidate A is known as an infrared source \citep{2006AJ....131.1163S}. Its WISE \citep{2010AJ....140.1868W} colors are similar with those of blazars and hence it is believed as a potential $\gamma$-ray emitter \citep{2012ApJ...748...68D}. RGs are also known as prominent extragalactic $\gamma$-ray emitters \citep{2010Sci...328..725A,2010ApJ...720..912A}. Though their Doppler boost effect is mild due to the large jet inclination angles, there are several dozens such sources detected by Fermi-LAT \citep{2022ApJS..260...53A}. Adopting FIRST 1.4~GHz \citep{2015ApJ...801...26H} and SDSS DR16 \citep{2020ApJS..249....3A} $g$-band measurements, its radio loudness $\mathcal{R}$ is yielded as $\sim$ 20.  However, such a relatively low radio loudness value does not imply a weak jet, since optical emissions of nearby RGs are typically dominated by the host galaxy components, rather than the accretion disk emissions. Interestingly, it is noted that a RLNLS1, SDSS J095909.51+460014.3 (hereinafter candidate B, $z$ = 0.399, \citealt{2017ApJS..229...39R}), also falls into the $\gamma$-ray localization uncertainty area. For candidate B, $\mathcal{R}$ is estimated as high as $\sim$ 1000, which is rather rare among NLS1s. Since the latter is not detected in \emph{WISE} \emph{W3} and \emph{W4} bands, the color comparison with blazars is unavailable. In this paper, considering the ambiguous association relationship, we perform thorough investigations on 4FGL 0959.6+4606 as well as its two potential counterparts, especially in the temporal perspective, attempting to straighten out the tangle. Here we adopt a ${\Lambda}$CDM cosmology with $\Omega_{M}$\,=\,0.32, $\Omega_{\Lambda}$\,=\,0.68,  and a Hubble  constant of $H_{0}$\,=\,67\,km$^{-1}$\,s$^{-1}$\,Mpc$^{-1}$ \citep{2014A&A...571A..16P}.

\section{Data reduction and analysis} \label{sec:data}

\subsection{{\it Fermi}-LAT data}
The first 13.8-yr (i.e. 2008 August 4 to 2022 June 4) Fermi-LAT {\tt Pass} 8 data ({\tt evclass} = 128 and {\tt evtype} = 3) are collected, with energy range from 100~MeV to 500~GeV. The {\tt Fermitools} software (version {\tt 2.0.8}), along with {\tt Fermitools-data} (version {\tt 0.18}), are adopted. In order to filter the photon data, the zenith angle cut (i.e. $< 90^{\circ}$) as well as the quality-filter cuts (i.e. {\tt DATA\_QUAL>0 \&\& LAT\_CONFIG==1}) are applied in the {\tt gtselect} and {\tt gtmktime} tasks. {\tt gtlike} task with {\tt Unbinned} likelihood approach is used to extract $\gamma$-ray flux and spectrum. Test statistic (TS = 2$\Delta\log\zeta$, \citealt{1996ApJ...461..396M}), where $\zeta$ represents maximum likelihood values of different models with and without the target source, is used to qualify the significance of $\gamma$-ray detection. During the likelihood analysis, diffuse $\gamma$-ray emission templates (i.e. {\tt gll\_iem\_v07.fits} and {\tt iso\_P8R3\_SOURCE\_V3\_v1.txt}) together with 4FGL-DR3 sources \citep{2022ApJS..260...53A} within 15$\degr$ of 4FGL 0959.6+4606 are considered. Parameters of the inner region background sources, a 10$\degr$ radius region of interest (ROI), as well as normalizations of the two diffuse templates are left free, while others are fixed in the 4FGL-DR3 values. We check whether there are new $\gamma$-ray sources based on the subsequently generated residual TS maps. If so, the likelihood fitting is then re-performed adopting the updated background model. When extracting $\gamma$-ray light curve, weak background sources (i.e. TS $<$ 10) are removed from the analysis model. 
\begin{figure}[t]
  \centering
    \includegraphics[width =9cm]{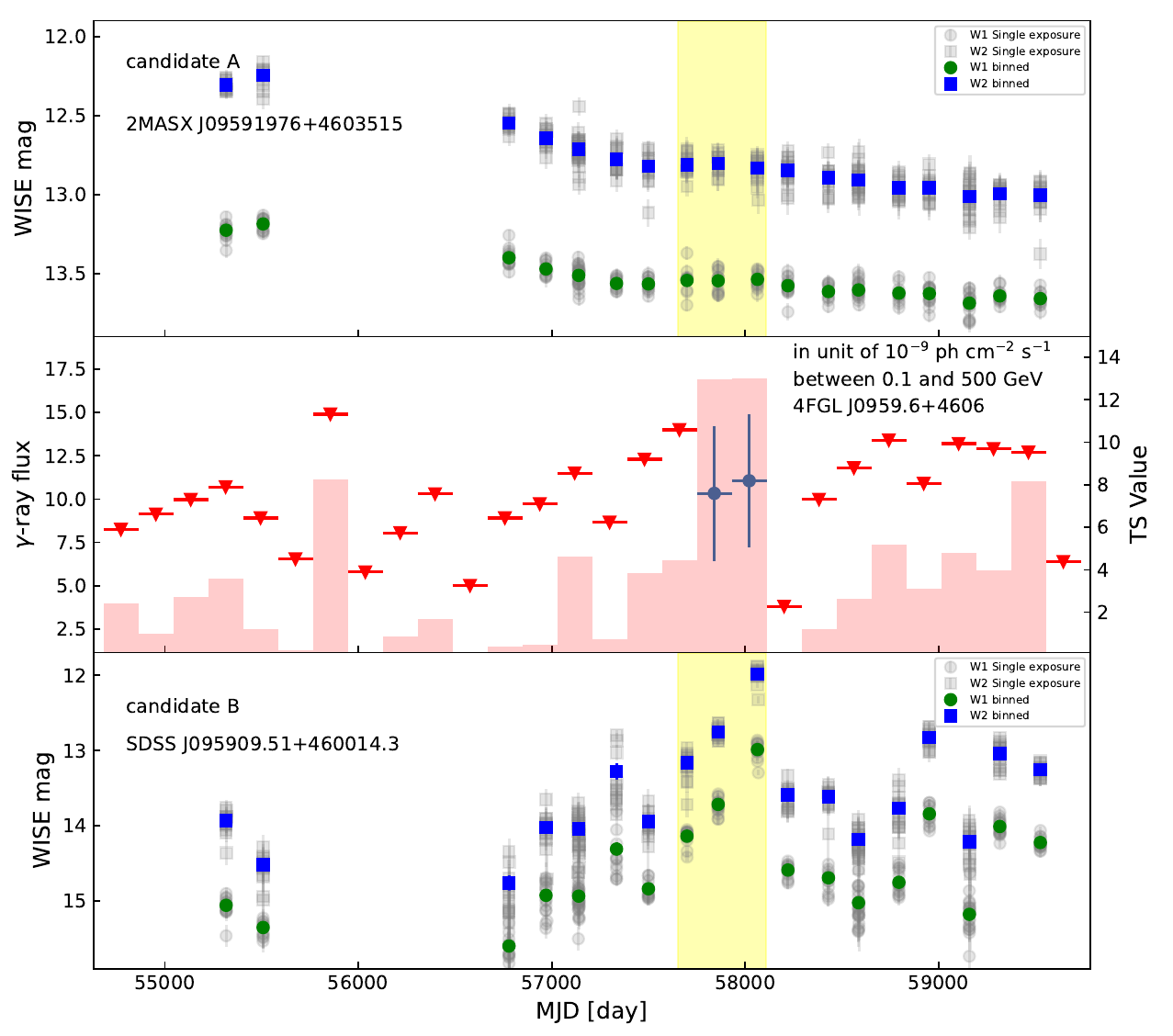}
    \caption{ $\gamma$-ray half-year time bin light curve of 4FGL J0959.6+4606, and infrared light curves of candidate A and candidate B. Blue circles represent the $\gamma$-ray fluxes, while the red triangles are upper limits. Red bars are the corresponding TS values. Yellow region marks the high flux state of 4FGL J0959.6+4606, identified by further monthly time bin light curve, see Figure \ref{gmlc}.}
   
    \label{mlc}
\end{figure}

\subsection{{\it Swift} data}
There is one visit with an exposure time of 1.9 ks  from the Neil Gehrels {\it Swift} Observatory \citep{2004ApJ...611.1005G} on candidate B at 8th May 2017. The data are analyzed by the FTOOLS software (version 6.28). For the XRT photon-counting mode data, firstly, event cleaning {\tt xrtpipeline} procedure with standard quality cuts is performed. Only 12 net photons of the target have been detected. Nevertheless, signal to noise ratio (i.e. S/N) of detecting the X-ray source is given as 3.4 by the {\tt ximage} task. No significant excess beyond the background at the position of candidate A is found. The spectrum of source is extracted from a circular region with a radius of 12 pixels, that of background from a circular blank region with a larger radius (50 pixels). The {\tt xrtmkarf} task is adopted to create the ancillary response files with the most recent calibration database. Considering the limited statistic, the data is not binned and we freeze the absorption column density to the Galactic value (i.e. $\rm 3.6\times10^{20}$ $\rm cm^{-2}$, \citealt{2016A&A...594A.116H}). Meanwhile, the photon index is set as a routine value (i.e. $\Gamma_{x} = 1.5$, , where $\Gamma_{x}$ is photon index of the power-law function). An unabsorbed 0.3-10.0 keV flux of $\rm 3.6^{+2.2}_{-1.6}\times10^{-13}$ erg $\rm cm^{-2}$ $\rm s^{-1}$ ($\mathcal{C}$-Statistic/d.o.f,  9.6/11; \citealt{1979ApJ...228..939C}) is yielded. In addition, there is a snapshot in UVOT {\it uw2} band, from which the magnitude is extracted by the aperture photometry (i.e. the {\tt uvotsource} task). During the extraction, a 5$\arcsec$ circular aperture is selected for the target while a larger source-free region is for the background.

\subsection{{\it WISE} data}
We collect the multi-photometric monitoring data in \emph{W1} and \emph{W2} bands (centered at 3.4, 4.6 $\mu$m in the observational frame) from Wide-field Infrared Survey Explorer (\emph{WISE}; \citealt{2010AJ....140.1868W}) and the Near-Earth Object {\it WISE} Reactivation mission (\emph{NEOWISE-R}; \citealt{2014ApJ...792...30M,neowise}). We remove the bad data points with poor image quality ("qi\_fact"$<$1), a small separation to South Atlantic Anomaly ("SAA"$<$5) and in the moon mask areas ("moon mask"=1) \citep{2017ApJ...846L...7S,2020ApJ...889...46S}. We bin the data in each epoch (nearly half year) using weighted mean value to probe the long-term variability of the target, the corresponding uncertainty is calculated by the propagation of measurement errors.

 \begin{table*}
\caption{The VLASS and FIRST frequencies, angular resolutions, radio integrated and peak flux densities, and background noise.}
\label{radio}
\centering
\begin{tabular}{cccccccc}
\hline\hline
\multirow{2}{*}{Name} & \multirow{2}{*}{Survey} & Frequency & Resolution & & $S_{int}$ & $S_{p}$ & RMS \\
& & (GHz) & (arcsec) & & (mJy) & (mJy beam$^{-1}$) & (mJy beam$^{-1}$) \\
\hline
\multirow{3}{*}{candidate A} & \multirow{2}{*}{VLASS} & \multirow{2}{*}{3.0} & \multirow{2}{*}{2.5} & Epoch 1 & 2.34 $\pm$ 0.3 & 1.67 $\pm$ 0.15 & 0.142 \\
& & & & Epoch 2 & 2.34 $\pm$ 0.18 & 1.875 $\pm$ 0.088 & 0.123 \\
& FIRST & 1.4 & 4.5 & & 4.06 & 3.81 & 0.137 \\
\hline
\multirow{3}{*}{candidate B} & \multirow{2}{*}{VLASS} & \multirow{2}{*}{3.0} & \multirow{2}{*}{2.5} & Epoch 1 & 28.5 $\pm$ 2.7 & 27.8 $\pm$ 1.5 & 0.203 \\
& & & & Epoch 2 & 32.36 $\pm$ 0.40 & 30.74 $\pm$ 0.22 & 0.126 \\
& FIRST & 1.4 & 4.5 & & 27.09 & 25.95 & 0.138 \\
\hline
\end{tabular}
% \tablefoot{}
\end{table*}

\subsection{Radio data}
  \begin{figure}
  \centering
    \includegraphics[width =9cm]{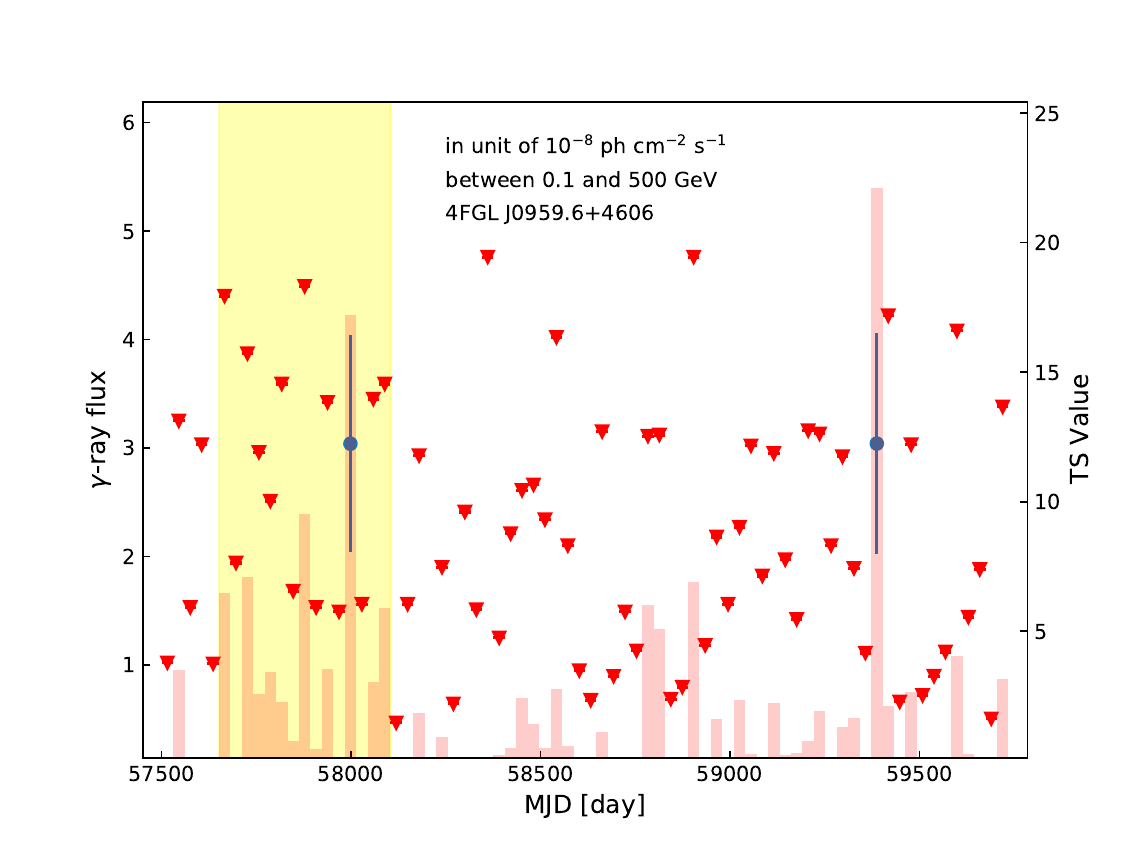}
    \caption{Monthly time bin $\gamma$-ray light curve of 4FGL J0959.6+4606, focusing the epoch of $>$ MJD 57500. The yellow region, totally 15-month, represents the high $\gamma$-ray flux state.}
    \label{gmlc}
 \end{figure}
We further cross-match the archive of the Very Large Array Sky Survey \citep[VLASS;][]{2020PASP..132c5001L} and the Faint Images of the Radio Sky at Twenty-Centimeters \citep[FIRST;][]{2015ApJ...801...26H} with optical positions of the two objects, candidate A and candidate B. The VLASS and FIRST frequencies, angular resolutions, radio integrated and peak flux densities, and background noise are listed in Table~\ref{radio}. The VLASS fluxes are obtained by modeling the source with a Gaussian fit on the image plane using the Common Astronomy Software Applications.

\section{Result} \label{sec:result}
\subsection{$\gamma$-ray and infrared behaviors}
Analysis of the entire 13.8-yr Fermi-LAT data confirms that there is a significant (TS = 89) $\gamma$-ray source in this direction. The corresponding powerlaw spectrum index is given as $\Gamma_{\gamma}$ = 2.61 $\pm$ 0.14, consistent with the values listed in 4FGL-DR3 \citep{2022ApJS..260...53A}. The optimized $\gamma$-ray location is at R.A. 149.866$\degr$ and DEC. 46.011$\degr$, with a 95\% confidence level (C. L.) error radius of 4.6$\arcmin$. Because of the proximity (roughly 4$\arcmin$), both low-energy sources remain within the $\gamma$-ray localization error radius.

Since the spatial resolution of \emph{W1} and \emph{W2} bands of \emph{WISE} is about 6$\arcsec$, the separated two infrared light curves of the counterparts provide crucial information of their temporal properties. As shown in Figure \ref{mlc}, variability amplitude of candidate A is about $\lesssim$ 0.5 mag, however, long-term variability with a brightening of $\sim$ 2.5 mag is seen in candidate B. Meanwhile, it is worth noting that the time epochs corresponding to the high flux states of the two sources are different. For candidate A, departed from the high flux state at the beginning of the \emph{WISE} observations around MJD 55500, a long flux decline until MJD 57500 is detected. Since then, its infrared fluxes maintains to be quiescent. For the RLNLS1, at the start of the \emph{WISE} observations (i.e. $<$ MJD 57500), no significant variations are detected. However, a distinct flux increase is then followed, peaking on MJD 58063. Other activities are also detected around MJD 59000. 

A half-year bin $\gamma$-ray light curve is extracted to compare with those in infrared bands, see Figure \ref{mlc}. Although the statistic is limited, two bins with TS values larger than 10 stand out. There is no corresponding infrared activity of candidate A matching the $\gamma$-ray brightening. However, the epoch of $\gamma$-ray flux increase coincides with the time of high flux state of infrared emission of candidate B, which is supported by the public yearly bin $\gamma$-ray light curve provided in the 4FGL-DR3 catalog \citep{2022ApJS..260...53A}. In addition, temporal properties of strong nearby $\gamma$-ray background sources are investigated, from which no similar behaviors to that of 4FGL 0959.6+4606 are found. Therefore, the brightening is likely intrinsic rather than artificially caused by the background sources. 

\begin{figure}
  \centering
    \includegraphics[width =9cm]{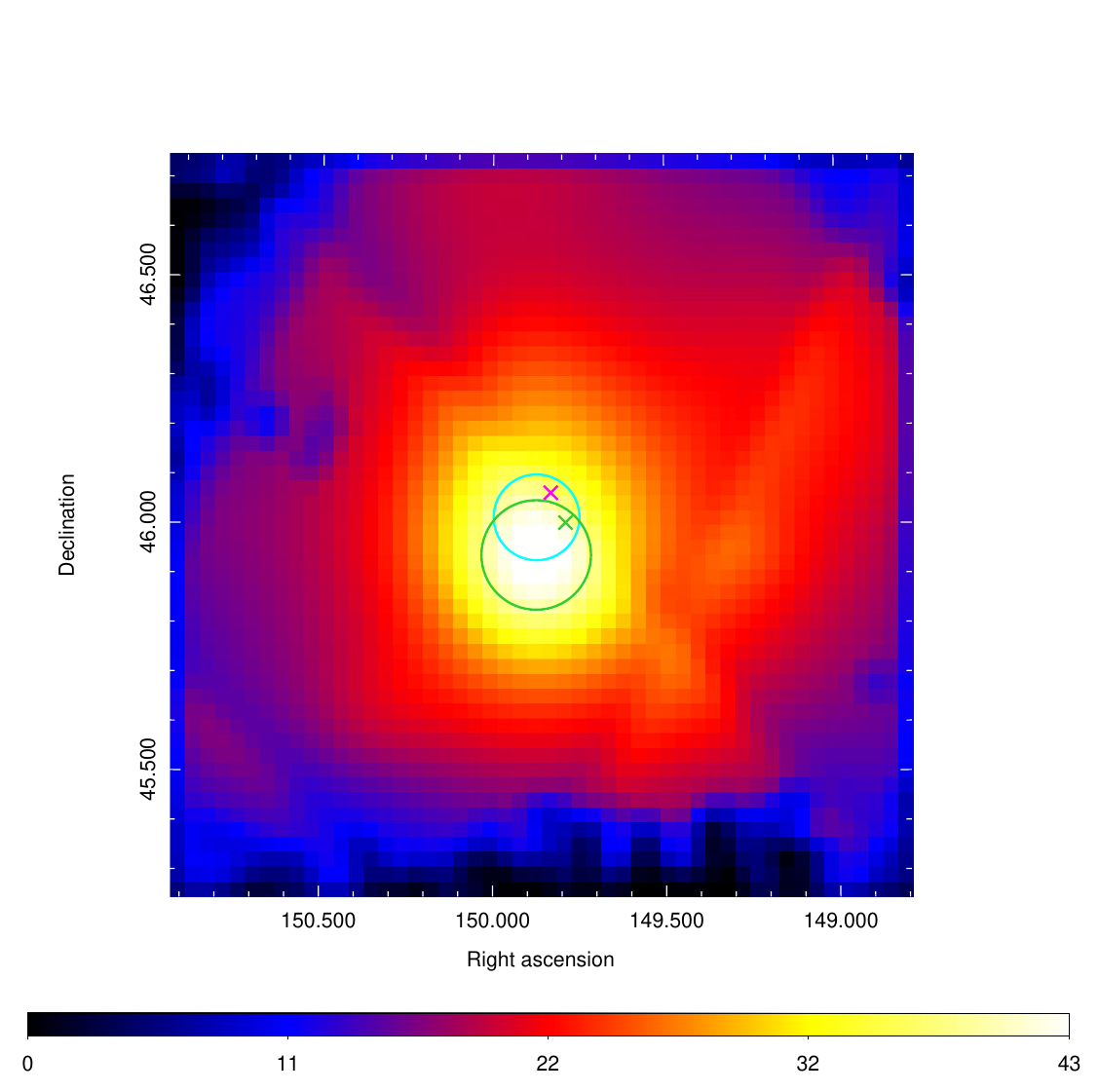}
    \caption{Smoothed $\gamma$-ray residual (i.e. 4FGL 0959.6+4606 is not included in the analysis model file) TS map with a scale of 1.5$\degr \times$1.5$\degr$ and 0.03$\degr$ per pixel, extracted from the 15-month Fermi-LAT data. The green and cyan circles correspond to the 95\% C.L. $\gamma$-ray localization uncertainty region of the 15-month and entire data, respectively. The green and pink X-shaped marker are radio positions of candidate B and candidate A, respectively}
    \label{tsmap}
\end{figure}

A monthly bin $\gamma$-ray light curve, focusing on the data later than MJD 57500, is also extracted to search potential short-term variation as well as determine the time range of the high flux state. Two time bins with relatively large TS values appear, locating at MJD 58000 and 59388, respectively. Interestingly, there are a few time bins with TS values $\ge$ 5 clustered around the former one. Therefore, an individual analysis, selecting a time range of 15-month Fermi-LAT data, has been performed, see Figure \ref{gmlc}. The analysis yields a significant (TS = 43) $\gamma$-ray source, of which an optimized $\gamma$-ray location at R.A. 149.875$\degr$ and DEC. 45.938$\degr$, along with a 95\% C. L. error radius of 6.6$\arcmin$, are derived. A corresponding TS map is generated, where localization results of data from different time epochs are shown together, see Figure \ref{tsmap}. In this case, the angular separation between the $\gamma$-ray location and it of candidate A is 7.8$\arcmin$, while the value for candidate B is 5.3$\arcmin$. Therefore, only the RLNLS1 falls into the $\gamma$-ray localization error radius. All the evidences suggest that candidate B is the low-energy counterpart of the $\gamma$-ray source. The photon flux of the $\gamma$-ray source is given as (1.40 $\pm$ 0.40) $\times$ $10^{-8}$ ph $\rm cm^{-2}$ $\rm s^{-1}$ and the $\Gamma_{\gamma}$ is constrained as 2.74 $\pm$ 0.21. Adopting a redshift of 0.399, a corresponding apparent luminosity of (3.1 $\pm$ 0.7) $\times$ $10^{45}$ erg $\rm s^{-1}$ is obtained. For the other monthly time bin at MJD 59388, further 10-day time bin light curve is extracted. In the time range between MJD 59369 and 59399, a $\gamma$-ray source (TS = 30) with a relatively hard spectrum ($\Gamma_{\gamma}$ = 1.90 $\pm$ 0.26) has been found, indicative of possible variability on timescale of a dozens of days. However, due to the limited statistics, in this case, both low-energy sources are embraced within the $\gamma$-ray localization uncertainty area. Moreover, no simultaneous infrared observations at this epoch are available.

\begin{figure}
  \centering
    \includegraphics[width =9cm]{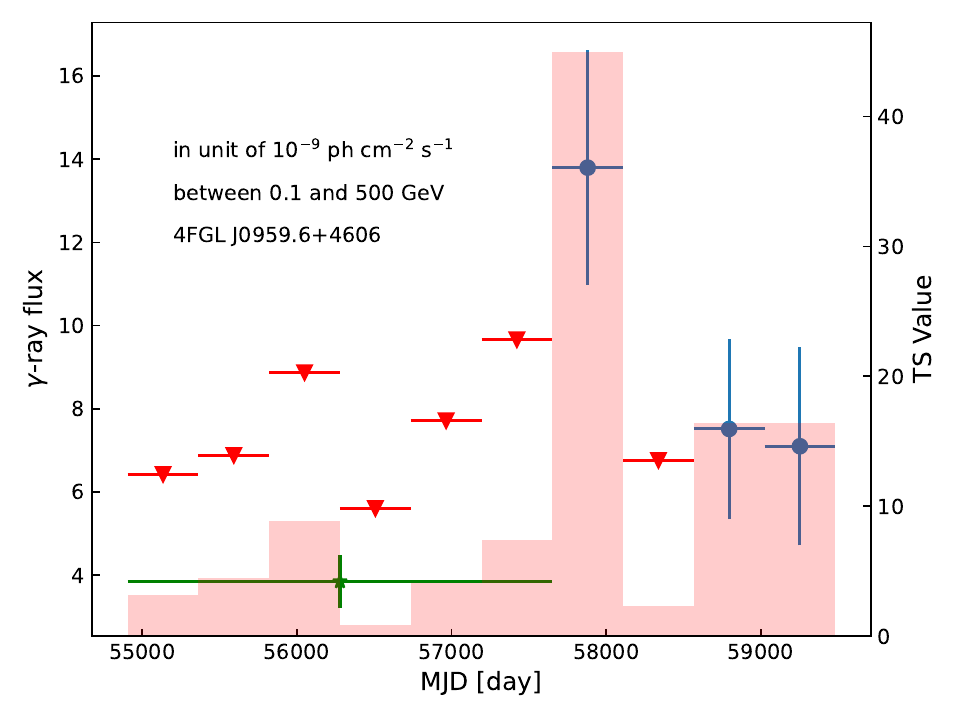}
    \caption{15-month time bin $\gamma$-ray light curve of 4FGL J0959.6+4606. Blue circles represent the $\gamma$-ray fluxes, while the red triangles are upper limits. Red bars are the corresponding TS values. The 7th bin corresponds to the yellow region in Figure \ref{gmlc}. The green star is the $\gamma$-ray flux from the first 7.5-yr data. }
     \label{15mlc}
\end{figure}

To further investigate the validness of the $\gamma$-ray brightening, a 15-month time bin light curve is also extracted, see Figure \ref{15mlc}, in which the 7th bin corresponds to the high flux state mentioned in Figure \ref{gmlc} and \ref{tsmap}. TS values are lower than 10 (i.e. $ < 2\sigma$) for the first six time bins in the light curve. However, TS value of the 7th bin reaches to 43 (i.e. 5.9$\sigma$). After the trial factor correction, the detection significance holds at 5.5$\sigma$, suggesting that the brightening is unlikely due to fluctuation of the background emissions. The significance of the flux enhancement is also looked into. Since the detection sensitivity of Fermi-LAT is strongly dependent on length of the exposure time and the target maintains at the low flux state at the beginning, analysis of the entire 7.5-yr data, corresponding to the first six time bins in the light curve, has been carried out. The analysis yields a marginal detection (i.e. TS =24), with a photon flux of  (3.84 $\pm$ 0.64) $\times$ $10^{-9}$ ph $\rm cm^{-2}$ $\rm s^{-1}$. A quantity defined as, $\sigma_{var} = \Delta flux/\sqrt{(fluxerr^{2}_{1}+fluxerr^{2}_{2})}$, is used to qualify the significance of variation between two different flux states. Hence the significance is calculated as 2.5$\sigma$, indicating that probability of the $\gamma$-ray flux enhancement is roughly 99\%. On the other hand,  amplitude of infrared variability is large ($\Delta mag >$ 2.5 mag), and the corresponding measurement uncertainties are relatively small (typically less than 0.1 mag), compared with that in $\gamma$ rays. More importantly, the two infrared light curves exhibit the similar variability trends, suggesting that infrared brightening is not caused by random fluctuation.

Beside the long-term variability, the single \emph{WISE} snapshots, typically a dozen of exposures within two days, allow us to investigate intraday variability. Assuming a Gaussian noise mode of WISE measurements \citep{2010AJ....140.1868W}, a $\chi^2$-test is adopted, of which the null hypothesis corresponds to a constant flux level \citep{2021ApJS..255...10M}. The value of the constant flux is optimized for each epochs to find a minimum reduced $\chi^2$ value. If such a value is still too large that the null hypothesis is rejected, the variability is decided to be significant. In addition to the estimated PSF photometry errors, systematic errors that are derived from the dispersion of the detections of the standard stars (i.e. $\sim 0.03$ mag, \citealt{2012ApJ...759L..31J}) have been also considered. No signs of significant infrared short-term variability (i.e. $< 3\sigma$) of candidate A are found. For candidate B, significant short-term variability (i.e. > 5$\sigma$) both in \emph{W1} and \emph{W2} bands around MJD 57334 and 58063 has been detected. Note that the latter is the time of maximum infrared flux value. Variability timescales in the source frame then are estimated as $\tau_{source}=\Delta t\times {\rm ln}2/{\rm ln}(F_{1}/F_{2})/(1+z)$, which suggest doubling timescales of $\simeq$ 9-hr and 17-hr, respectively. 

\subsection{Re-analysis of the SDSS spectrum}
\begin{figure}
  \centering
    \includegraphics[width =8cm]{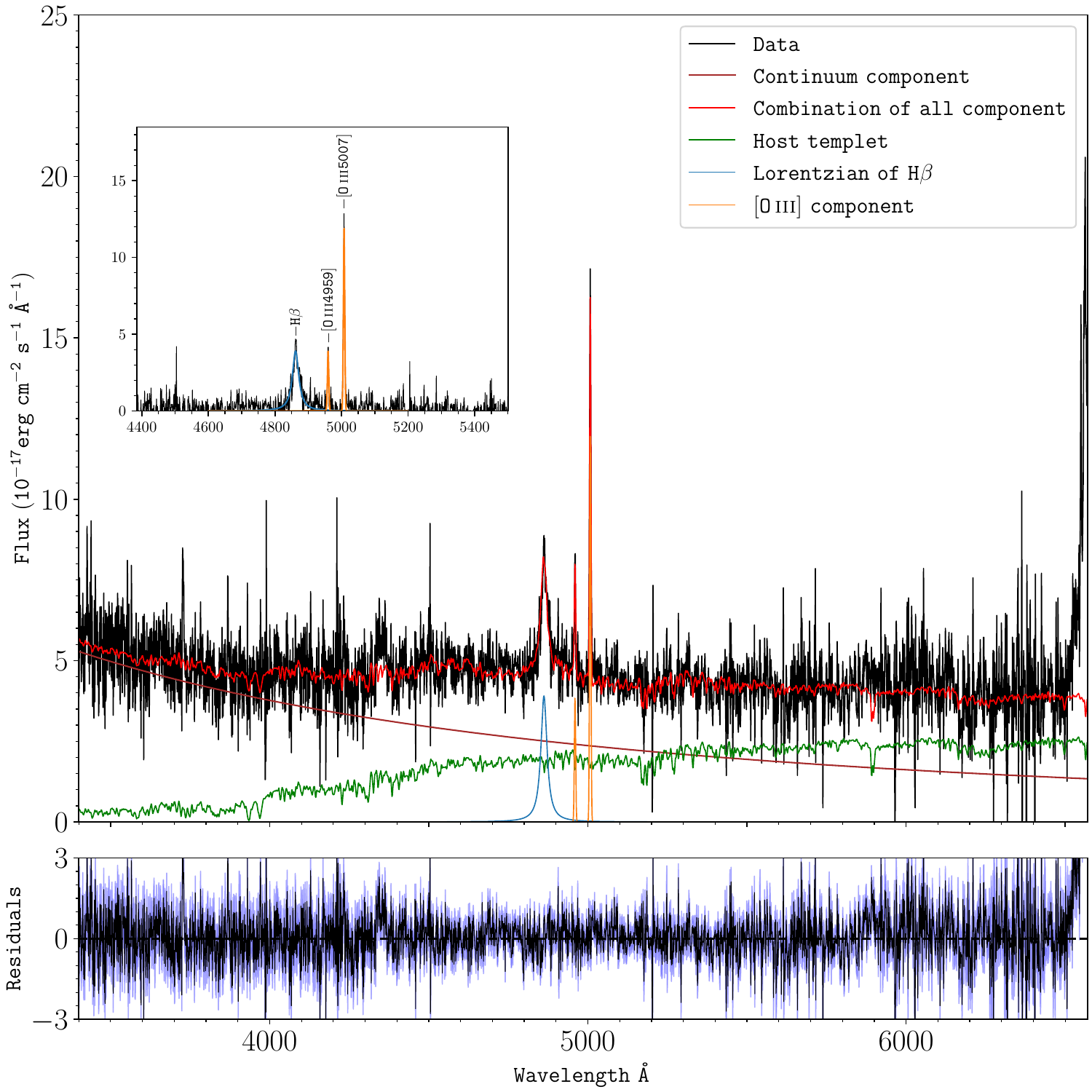}
    \caption{The SDSS spectrum of candidate B. The red line corresponds to the combination of all components. The green, brown, blue and orange lines represent the descriptions of host galaxy, power-law continuum, Lorentzian profile of H$\beta$ as well as the [O\,\textsc{iii}] doublet, respectively.}
    \label{sdss}
\end{figure}

The optical spectrum of candidate B observed on MJD 54525 is derived from the SDSS archive, and its median S/N is about 8. The analysis is with the help of {\tt PyQSOFit} software\footnote{https://github.com/legolason/PyQSOFit} \citep{2018ascl.soft09008G}. After shifting the spectrum to its rest frame and considering the Galactic extinction \citep{2011ApJ...737..103S}, firstly, it is simultaneously fitted with a global AGN power-law continuum and stellar contribution of host galaxy while emission lines excluding the Fe\,\textsc{ii} multiplets \citep{1992ApJS...80..109B,2001ApJS..134....1V} are masked. The galaxy emission is described by summation of independent component analysis templates \citep{2006AJ....131..790L}. Then a simultaneous fitting of the H$\beta$ emission line region (i.e. 4385–5500\AA) is carried out, in which the local continuum, the Fe\,\textsc{ii} multiplets together with the H$\beta$ and the [O\,\textsc{iii}] doublet emission lines are considered, see Figure \ref{sdss}. However, strength of the Fe\,\textsc{ii} multiplets can not be well constrained due to the relatively low S/N. He\,\textsc{ii} $\lambda$4687\AA~line also falls into this region but is not distinct from the continuum emission. Monte-Carlo (MC) simulations are performed to estimate the measurement uncertainties. Each data point is treated as a gaussian probability distribution, of which the mean and standard deviation are set as the measured value of the data point and its error. Then 1000 times MC samplings for each data point bring mock spectra. By repeatedly fitting the mock spectra, distributions of the parameters are obtained. The measurement uncertainties are calculated based on from the 68\% range (centered on the median) of the distributions \citep{2011ApJS..194...45S}. A single Lorentzian profile gives an acceptable description of the H$\beta$ line, of which the full width at half maximum (FWHM) is obtained as (1424 $\pm$ 180) km/s. A two Gaussian model does not bring significant improvement of the fitting. In addition, flux ratio of total [O\,\textsc{iii}] to total H$\beta$ is estimated as 0.7. Therefore, candidate B is confirmed as a NLS1, consistent with the results of \cite{2017ApJS..229...39R}. Using the empirical virial BH mass relation \citep{2008ApJ...680..169S} and the estimated continuum luminosity $\rm \lambda \rm L_{\lambda}$(5100 \AA) = (6.45 $\pm$ 2.76)$\times$ $10^{43}$ erg $\rm s^{-1}$, we get a constraint of $\rm log(M_{BH}/M_{\sun}) \gtrsim$ 6.8, in case of a disk-like geometry of the broad line region (BLR). Moreover, the Eddington ratio $L_{bol}/L_{edd}\lesssim0.6$ can be inferred, using the bolometric luminosity correction $L_{bol}\sim(8.1\pm0.4)L_{5100 \AA}$ \citep{2012MNRAS.422..478R}.
% 1,000 mock spectra by adding Gaussian noise to the original spectrum are fitted repeatedly.}

\subsection{Radio properties}
We measure the spectral index between 1.4 and 3.0 GHz by fitting the spectrum with a power-law using the definition
\begin{equation}
\alpha = \frac{\log(S_2/S_1)}{\log(\nu_2/\nu_1)}
\end{equation}
where S1 is the flux density at 1.4 GHz from FIRST and S2 is the average flux density at 3.0 GHz of the two epochs VLASS observations. Candidate A has a steep spectral slope with $\alpha = -0.72$ and candidate B has a flat spectral slope with $\alpha = 0.15$. The two sources show a point source morphology in the VLASS and FIRST surveys. The ratio of peak/integrated flux density of candidate A is smaller than that of candidate B, which indicates that the former one is more compact.
%Both the spectral slope and the compactness suggest that candidate B is more likely to produce a jet than candidate A.

\subsection{Implications of broadband SED of the RLNLS1}
\begin{figure}
  \centering
    \includegraphics[width =8cm]{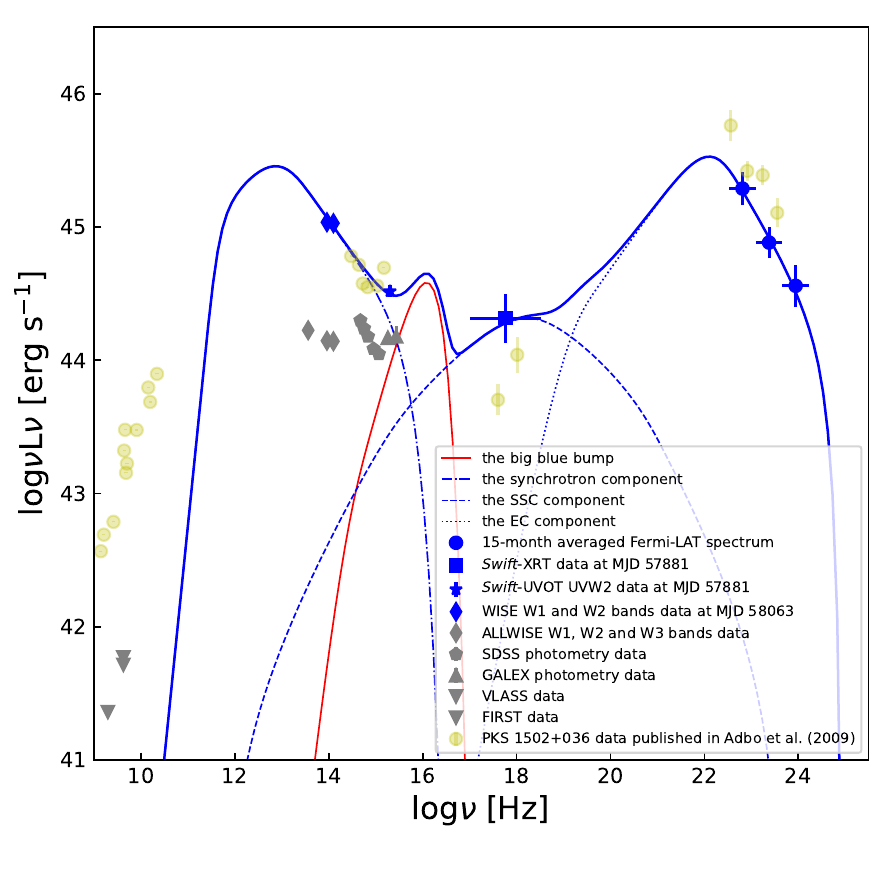}
    \caption{SED in high flux state of SDSS J095909.51+460014.3 along with the theoretical descriptions. Archival SED data of PKS 1502+036 \citep{2009ApJ...707L.142A} have been also plotted for comparison. Un-simultaneous data are colored in grey. The red line represents the description of the big blue bump,  with $L_{d} = 8\times 10^{44}$ erg $\rm s^{-1}$ ($\sim 0.4L_{edd}$) and $M_{BH} = 6.8\times 10^{6}M_{\odot}$.}
    \label{sed}
\end{figure}

%\subsection{candidate B as a new $\gamma$-NLS1}
Broadband SED of candidate B has been drawn based on the multiwavelength observations, especially focusing on the simultaneous brightening of the $\gamma$-ray and infrared emissions. The high flux state SED includes the 15-month $\gamma$-ray spectrum between MJD 57651 and 58104, {\it Swift} observations at MJD 57881, as well as \emph{WISE} detections at MJD 58063, see Figure \ref{sed}. Un-simultaneous data including archival GALEX \citep{2012AAS...21934001S}, SDSS \citep{2020ApJS..249....3A} and ALLWISE \citep{allwise} photometric data, as well as the VLASS \citep{2020PASP..132c5001L} and FIRST \citep{2015ApJ...801...26H} radio data, are also plotted. Considering it shares a similar redshift with PKS 1502+036, data of the latter from \cite{2009ApJ...707L.142A} are drawn for comparison. A remarkable feature of the SED of candidate B is violent variation in the infrared band, which makes the jet emission is overwhelming even in the UV domain then. Shape of the SED in optical-UV bands of candidate B is similar to that of PKS 1502+036, suggesting that the peak frequencies of their synchrotron bump are likely $\lesssim$ $\rm 10^{14}$ Hz. Therefore, candidate B is classified as a low-synchrotron-peaked source (LSP; \citealt{1995ApJ...444..567P,2010ApJ...716...30A}). For the high energy bump, both of the sources exhibit soft $\gamma$-ray spectra, though PKS 1502+036 is more luminous than candidate B. 

\begin{table}
\caption{List of input-parameters adopted in the SED modelling, also see Figure \ref{sed}.}
\centering
\begin{tabular}{cccccccccc}
\hline\hline
 K$(cm^{-3})$ & $\gamma_{br}$ & $p_1$ & $p_{2}$  & B(gauss) & $\delta$ &R$'_j$(cm)\\
\hline
8.7$\times 10^3$ &357 & 2.0 & 4.0 & 1.1 &13.5 & 2.5$\times 10^{16}$\\
\hline
\end{tabular}
\tablefoot{K is the normalization of the particle number density; $\gamma_{br}$ is the break energy of the electron distribution ; $p_{1,2}$ are the indexes of the broken power-law radiative electron distribution; B is the magnetic field strength; $\delta$ is the Doppler boosting factor; R$'_j$ is the radius of the emission blob in the jet comoving frame. The minimum energy of the electrons is set as 10, while the corresponding maximum energy is set as 20 times of the $\gamma_{br}$. }
\end{table}

The intraday infrared variability allow us to put a constraint of Doppler factor (i.e. $\delta$) of the jet. It should be large enough to avoid severe attenuation on $\gamma$-rays from soft photons via the $\gamma\gamma$ process. The corresponding opacity \citep{1995MNRAS.273..583D} is given as: 
\begin{equation}
         \tau_{\gamma\gamma}(x^{\prime})=\frac{\sigma_{\rm T}}{5}n^{\prime}(x^{\prime}_{\rm t})x^{\prime}_{\rm t}R^{\prime},
\end{equation}
where $\sigma_{\rm T}$ is the Thomson scattering cross-section, $R^{\prime}$ is the absorption length, $x^{\prime}$ = $h\nu^{\prime}/m_{e}c^{2}$ and $x^{\prime}_{\rm t}$ correspond to the dimensionless energy of the $\gamma$ rays and target soft photon in the comoving frame, $n^{\prime}(x^{\prime}_{\rm t})$ is the differential number density per energy of the latter. The target photons from the jet itself could be responsible for the absorption. Hence radius of the jet dissipation region, constrained as $R_{j}^{\prime}\leq$ c$\tau_{source} \delta$, where $\tau_{source}$ is set as 17-hr, is equal to the absorption length. The most energetic photon associated with the source is about 4~GeV. Adopting the observed luminosity of soft photons at a few keV of 2 $\rm \times 10^{44}$ erg $\rm s^{-1}$, a constraint of $\delta \gtrsim$ 3 is given. Alternatively, lack of information about the external absorption photons at several tens of eV prevents us from setting a reliable constraint.

The high flux state SED is described by the classic single-zone homogeneous leptonic model including the synchrotron and IC processes, in which the synchrotron self-absorption process and the Klein$-$Nishina effect in the IC scattering are considered. A relativistic compact blob, embedded in the magnetic field and the external photon field, is responsible for the jet emission. Relativistic transformations of luminosity and frequency between the observational frame and the jet frame are $\nu L_{\nu} = \delta^{4}\nu^{\prime}L^{\prime}_{\nu^{\prime}}$ and $\nu = \delta\nu^{\prime}/(1+z)$, respectively. The emitting electrons are assumed to follow a broken power-law distribution,
\begin{equation}
N(\gamma ) \propto \left\{ \begin{array}{ll}
                    \gamma ^{-p_1}  &  \mbox{ $\gamma_{\rm min}\leq \gamma \leq \gamma_{br}$} \\
            \gamma _{\rm br}^{p_2-p_1} \gamma ^{-p_2}  &  \mbox{ $\gamma _{\rm br}<\gamma\leq\gamma_{\rm max}$, }
           \end{array}
       \right.
\label{Ngamma}
\end{equation}
where $p_{1,2}$ are indices of the broken power-law particle distribution, and $\rm \gamma_{br}$, $\rm \gamma _{min}$ and $\rm \gamma _{max}$ correspond to the break, the minimum and maximum energies of the electrons respectively. Since the accretion system of NLS1s is radiatively efficient, EC processes are proved to be crucial to describe the $\gamma$-ray emission of the RLNLS1s \citep{2009ApJ...699..976A,2009ApJ...707L.142A}. Distance (i.e. $D_{diss}$) between the jet dissipation region and the central SMBH is needed to determine the origin of the external soft photon field \citep{2016ARA&A..54..725M}. Assuming a simple conical jet geometry, the distance is in proportion to the radius of the jet blob $R_{j}^{\prime}$ \citep{2010MNRAS.405L..94T}, which can be constrained based on the WISE short-term variability. Taking a routine $\delta$ value (i.e. 15, \citealt{2017MNRAS.466.4625L}), the distance is suggested as $D_{diss}\simeq \delta R_{j}^{\prime} \sim 0.1$ pc. Meanwhile, the radius of BLR can be constrained $\sim 0.02$~pc based on the measured $\rm L_{\lambda} (5100 \AA)$ \citep{2013ApJ...767..149B}. Therefore, in EC process, infrared emissions from the dust torus with temperature of 1200 K \citep{2000ApJ...545..107B} are considered as the soft photon field, of which the energy density in the rest frame is set as 3 $\rm \times 10^{-4}$ erg $\rm cm^{-3}$ \citep{2012MNRAS.425.1371G}. In perspective of the broad emission line revealed by the SDSS spectrum, contribution of the accretion disk emission \citep{1973A&A....24..337S} has been considered. The accretion disk extends from 3 $\rm R_{s}$ to 2000 $\rm R_{s}$, where $\rm R_{s}$ is the Schwarzschild radius. It produces a total luminosity of $L_{d}=\eta\dot{M}c^{2}$, in which $\dot{M}$ is the accretion rate and $\eta$ is the accretion efficiency. The efficiency is set to 0.057 for a low angular momentum black hole \citep{2002apa..book.....F}. The accretion disk emission is characterized by a multi-temperature radial profile \citep{2002apa..book.....F}, with a local temperature T at a certain radius $R_{disk}$ is given by,
\begin{equation}
\rm T^{4} = \frac{3R_{s}L_{d}}{16\pi\eta\sigma_{SB}R_{disk}^{3}}\left[1-(\frac{3R_{s}}{R_{disk}})^{1/2}\right], 
\end{equation}   
where $\rm \sigma_{SB}$ is the Stefan–Boltzmann constant.

The theoretical description provides a well explanation of the high flux SED, shown in Figure \ref{sed}. Corresponding input parameters are listed in Table 2. Lack of sub-mm/far-infrared data as well as a reliable X-ray spectrum obstacle to a further constraint of the modeling. Meanwhile, the disk model can not be well constrained from the high flux SED that the contribution of jet is overwhelming in the UV domain then. Nevertheless, the disk model manages to provide an acceptable description of the archival GALEX data that the jet emission was in the quiescent state. Similar with the known $\gamma$-NLS1s \citep[e.g.,][]{2009ApJ...707L.142A},  $\gamma$-ray emission of candidate B is mainly from the EC processes. Furthermore, the input parameters here are consistent with that used in those sources \citep[e.g.,][]{2012MNRAS.426..317D,2013MNRAS.433..952D,2018MNRAS.477.5127Y,2021MNRAS.504L..22R,2021A&A...649A..77G}, such as B $\sim$ 1 Gauss, $\delta \sim 15$ and $\gamma_{br} \sim$ 500. Interestingly, the $\gamma_{br}$ values of $\gamma$-NLS1s, alike of FSRQs, is generally lower than those of BL Lacs \citep[e.g.,][]{2018ApJS..235...39C}. In fact, majority of $\gamma$-NLS1s, are LSPs \citep{2022ApJS..260...53A}. In addition to the leptonic scenarios, hadronic scenarios could also play an important role in $\gamma$-NLS1s, and hence they are possible neutrino emitters \citep{2015RPPh...78l6901A}. Recently, a cospatial incoming IceCube neutrino event is temporally coincident with a minor $\gamma$-ray flare of 1H 0323+342 \citep{2020ApJ...893..162F}. Hadronic processes have been taken into account to describe the SED of PKS 1502+036 \citep{2023ApJ...942...51W} We have looked up into the IceCube alert data archive \footnote{https://icecube.wisc.edu/science/real-time-alerts/}, unfortunately, no known neutrino event in the direction of candidate B is found. Moreover, no evidence of hadronic processes in the electromagnetic data, for example, an orphan $\gamma$-ray flare \citep[e.g.,][]{2013ApJ...768...54B}, has been found either. Future multi-messenger studies are needed to probe its potential as a neutrino source.

\section{Discussion and summary}
Since $\gamma$-NLS1s shed lights on jet launching from extreme environments, efforts of enlarging their number have been devoted. Majority of them are found by detections of new $\gamma$-ray sources towards to the RLNLS1s \citep[e.g.,][]{2009ApJ...707L.142A}. Re-analysing optical spectra of known $\gamma$-ray sources to identify their NLS1 nature is another approach \citep[e.g.,][]{2015MNRAS.454L..16Y}. In particular, infrared spectroscopic observations are needed for high-redshift ones (i.e. z $\gtrsim$ 0.8) since H$\beta$ line is then not embraced by the optical spectral coverage \citep{2019MNRAS.487L..40Y,2021MNRAS.504L..22R}. Here another approach is adopted. Endeavoring to pin down an association relationship between a known $\gamma$-ray source and a RLNLS1 so that a new $\gamma$-NLS1 can be identified. In fact, due to the proximity between candidate A and candidate B, and the weak spectrally soft $\gamma$-ray signal, the identification is impossible without the help of infrared and $\gamma$-ray temporal information. Considering candidate B as an LSP and the relatively soft $\gamma$-ray spectrum, energy regimes of these emissions locate at the right side of the synchrotron and IC SED bump peaks respectively, and hence they are likely from the same population of emitting electrons under the leptonic scenario. The common origin makes contemporaneous brightening of the infrared and gamma-ray emissions reasonable. A similar phenomenon has been reported for a high-redshift FSRQ \citep{2019ApJ...879L...9L}. On the other hand,  $\gamma$-ray variability behaviors of radio galaxies are diverse. No significant variations at a timescale of months are commonly observed\footnote{https://fermi.gsfc.nasa.gov/ssc/data/access/lat/12yr\_catalog/4FGL-DR3\_LcPlots\_v29.tgz}. However, occasionally, flare-like phenomena for some individuals have been reported. For instance, a multifrequency campaign of 3C 111 reveals simultaneous flux increases in the millimeter, optical and X-ray fluxes when its $\gamma$-ray emission is bright \citep{2012ApJ...751L...3G}. Multi-wavelength monitoring of 3C 120 detects correlated radio and $\gamma$-ray activities \citep{2015ApJ...799L..18T}.  Therefore, if candidate A is the low-energy counterpart, it is reasonable to assume that its $\gamma$-ray flux is proportional of the infrared flux. However, the $\gamma$-ray flux increase accompanying with its quiescent infrared emissions is not supportive for this assumption. There is no categorized $\gamma$-ray source in 3FGL that is built by the first 4-year Fermi-LAT data, when infrared flux of candidate A is in high flux state. The failure of a $\gamma$-ray detection then puts a severe limit of its expected flux level. Such an expectation at MJD > 57000 is even more lower due to its infrared flux decay. In the epoch of 15-month when the infrared flux of candidate B is brightening, the $\gamma$-ray contribution of candidate A is likely negligible. In conclusion, the RLNLS1 is preferred as the counterpart of 4FGL J0959.6+4606. Upcoming additional optical sky surveyors in time domain, like the Wide Field Survey Telescope \citep{2016SPIE10154E..2AL}, as well as the Large Synoptic Survey Telescope \citep{2019ApJ...873..111I}, will be crucial in investigation of multi-wavelength temporal properties of AGN jets.

\begin{figure}
  \centering
    \includegraphics[width =10cm]{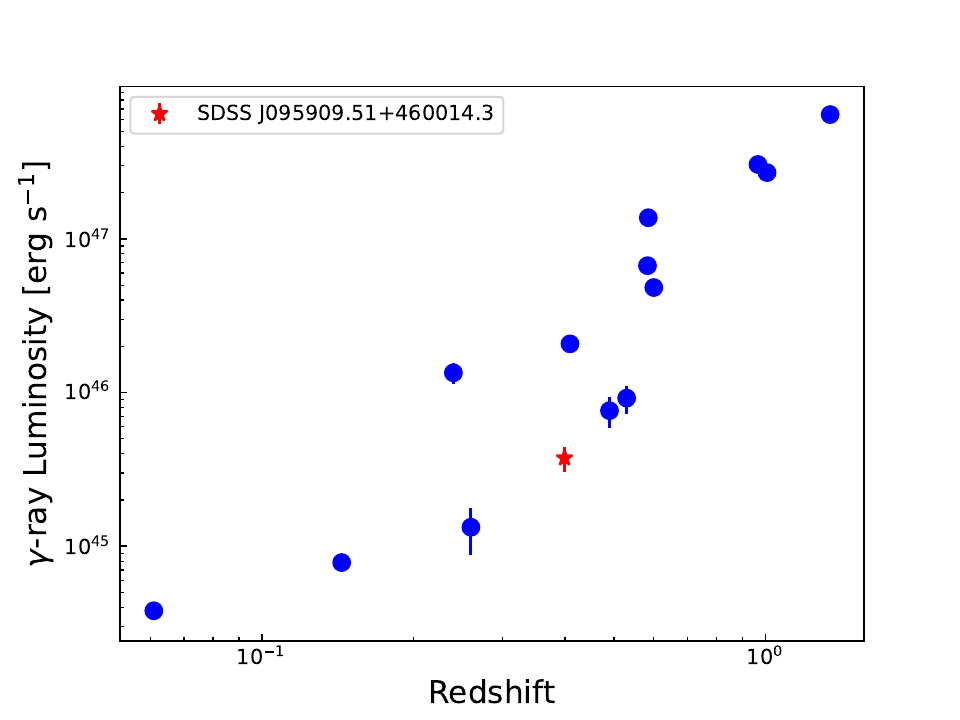}
    \caption{Comparison between SDSS J095909.51+460014.3 and other $\gamma$-NLS1s. K-corrections for the luminosities have been applied. Luminosity of the target is based on the analysis of the 15-month Fermi-LAT data. For other sources, each luminosity value is calculated from energy flux of the time bin with the highest flux level in public yearly bin light curve \citep{2022ApJS..260...53A}.}
    \label{comp}
\end{figure}

It is interesting to compare candidate B with the known highly-beamed $\gamma$-NLS1s. Firstly, $\gamma$-NLS1s tend to exhibit soft $\gamma$-ray spectra (i.e. $\Gamma_{\gamma} \gtrsim$ 2.5, \citealt{2022ApJS..260...53A}), consistent with that of candidate B. For comparison of the $\gamma$-ray luminosity, since a reliable estimation of candidate B is only derived in the high flux state (i.e. in a 15-month period), energy flux of the time bin with the highest flux level in public yearly bin light curve for each known $\gamma$-NLS1 has been extracted \citep{2022ApJS..260...53A}. As shown in Figure \ref{comp}, candidate B lies at the low luminosity region of the distribution. It is not surprising because candidate B is one of the few sources with $z \le$ 0.4. In the temporal perspective, detections of $\gamma$-ray flares are in common for these sources \citep[e.g.,][]{2016MNRAS.463.4469D,2018MNRAS.477.5127Y,2021A&A...649A..77G}. In an extreme case, the variability amplitude of PMN J0948+0022 can reach up to more than one order of magnitude \citep{2015MNRAS.446.2456D}. Moreover, simultaneous broadband activities of $\gamma$-NLS1s have been also observed \citep[e.g.,][]{2015MNRAS.446.2456D}, suggesting that their central engine resemble that of blazars. The relatively low luminosity compared with FSRQs, is likely due to the relatively lower black hole masses \citep{2011ApJ...735..108C}. 

Nearby $\gamma$-NLS1s are valuable targets for investigations on the host galaxy environments where the strong relativistic jets are launched. Based on the Hubble Space Telescope observations on 1H 0323+342, its host galaxy is suggested as a one-armed spiral galaxy \citep{2007ApJ...658L..13Z}. Alternatively, a ring structure likely due to a galaxy merger has been identified by NOT images \citep{2008A&A...490..583A,2014ApJ...795...58L}. The host galaxy of FBQS J1644+2619 is decomposed with the combination of a pseudobulge, a disc and a ring component indicative of a late-type galaxy \citep{2017MNRAS.467.3712O}, while a bulge component with a large S$\rm \acute{e}$rsic index (n = 3.7), typical for a elliptical galaxy, is found \citep{2017MNRAS.469L..11D}. In addition, studies on PKS 2004-447 as well as SDSS J211853.33-073214.3 reveal pseudobulges \citep{2016ApJ...832..157K,2020ApJ...892..133P}. In view of the controversy, investigation of host galaxy morphology of candidate B is helpful. Discrepancy of BH masses from different estimation approaches has been noticed \citep{2016MNRAS.458L..69B,2021A&A...654A.125B}. Sources with small inclination angles, like the $\gamma$-NLS1s, if they possess a flattened BLR, the BH masses induced by the emission line widths can be seriously underestimated. A reliable measurement of bulge luminosity of candidate B would set an independent constraint and is crucial for understanding its nature. Future optical spectropolarimetry observations will also help to distinguish these two scenarios.

In summary, we have performed thorough investigations on the $\gamma$-ray source 4FGL 0959.6+4606, as well as its two potential low-energy counterparts. They both fall into the $\gamma$-ray localization uncertainty area corresponding to the entire 13.8-yr Fermi-LAT data. Re-analysis of the SDSS spectrum of candidate B confirms that it is a NLS1. Infrared flux descents of 0.5 mag of candidate A are detected, while brightening of 2.5 mag are observed for candidate B. $\gamma$-ray light curves of 4FGL 0959.6+4606 reveal a flux enhancement in an epoch of 15-month. No infrared activities of candidate A are found then. However, a temporally coincident infrared brightening of candidate B appears.  An individual analysis of the Fermi-LAT data, representative of the high flux $\gamma$-ray state, yields a significant $\gamma$-ray source (TS = 43). At this time, only candidate B is embraced by the $\gamma$-ray localization uncertainty area. The association relationship between candidate B and 4FGL 0959.6+4606 is supported in respect of both temporal and spatial evidences. Therefore, we conclude that candidate B is likely a $\gamma$-NLS1. In addition to the $\gamma$-ray and infrared data, multiwavelength data are collected and analyzed to construct a broadband SED of the RLNLS1 for the high flux state. The single-zone homogeneous leptonic jet modeling provides an acceptable description of the SED. The jet properties of candidate B are found to be comparable with that of known $\gamma$-NLS1s. Future high resolution imaging observations will reveal its host galaxy morphology and revisit its BH mass estimation.

\begin{acknowledgements}
We appreciate the instructive suggestions from the anonymous referee. This research has made use of data obtained from the High Energy Astrophysics Science Archive Research Center (HEASARC), provided by $\rm NASA^{\prime}$s Goddard Space Flight Center. This research makes use of data products from the Wide-field Infrared Survey Explorer, which is a joint project of the University of California, Los Angeles, and the Jet Propulsion Laboratory/California Institute of Technology, funded by the National Aeronautics and Space Administration. This research also makes use of data products from NEOWISE-R, which is a project of the Jet Propulsion Laboratory/California Institute of Technology, funded by the Planetary Science Division of the National Aeronautics and Space Administration. 

This work was supported in part by the NSFC under grants 11703093, U2031120, 12103048, and 11833007, as well as the Project funded by China Postdoctoral Science Foundation: 2020M682013. This work was also supported in part by the Special Natural Science Fund of Guizhou University (grant No. 201911A) and the First-class Physics Promotion Programme (2019) of Guizhou University. S.C. acknowledges support by the Israel Science Foundation (grant no.1008/18) and a Center of Excellence of the Israel Science Foundation (grant no.2752/19).
\end{acknowledgements}

\bibliographystyle{aa} 
\bibliography{refs}

%-------------------------------------------------------------

\end{document}